\documentclass[aps,prb,superscriptaddress,showpacs,twocolumn,amsmath]{revtex4-1}
\usepackage{dcolumn}
\usepackage{multirow}
\usepackage{graphicx}
\usepackage{subfigure}
\usepackage{epsf}
\usepackage{epstopdf}
\usepackage{color}
\newcommand{\bk}{{\bf k}}
\newcommand{\bq}{{\bf q}}
\newcommand{\bK}{{\bf K}}
\newcommand{\bQ}{{\bf Q}}
\begin{document}
\title{Large phonon-drag enhancement induced by narrow quantum confinement at the LaAlO$_3$/SrTiO$_3$ interface}
\author{I. Pallecchi}
\author{F.Telesio}
\author{D. Marr\'e}
\affiliation{CNR-SPIN UOS Genova and Dipartimento di Fisica, Via Dodecaneso 33, 16146 Genova, Italy}
\author{D. Li}
\author{S. Gariglio}
\author{J.-M. Triscone}
\affiliation{DQMP, University of Geneva, 24 Quai Ernest-Ansermet, 1211 Geneva 4, 
Switzerland}
\author{A. Filippetti}
\affiliation{CNR-IOM UOS Cagliari, S.P. Monserrato-Sestu Km.0,700, Monserrato (Ca), 09042, Italy}

\date{\today}
\begin{abstract}
The thermoelectric power of the two-dimensional electron system (2DES) at the LaAlO$_3$/SrTiO$_3$ interface is explored below room temperature, in comparison with that of Nb-doped SrTiO$_3$ single crystals. For the interface we find a region below $T$=50 K where thermopower is dominated by phonon-drag, whose amplitude is hugely amplified with respect to the corresponding bulk value, reaching values $\sim$mV/K and above. The phonon-drag enhancement at the interface is traced back to the tight carrier confinement of the 2DES, and represents a sharp signature of strong electron-acoustic phonon coupling at the interface.
\end{abstract}

\pacs{}

\maketitle
\section{Introduction}

The rise of thermoelectric efficiency in low-dimensional materials\cite{Hicks1993,Hicks1993a,Hicks1993b,Broido1995,Hicks1996,Mahan1996,Dresselhaus2007,Sofo1994} is a long-standing source of inspiration for material design\cite{Venka2001,Majumdar2004,Snyder2008,Biswas2012,Zebarjadi2012,Cahill2014,Beekman2015}. The potential of LaAlO$_3$/SrTiO$_3$\cite{Ohtomo2004,Huijben2006,Siemons2007,Nakagawa2006,Reyren2007,Mannhart2010} and, more in general, of SrTiO$_3$-based heterostructures as thermoelectric materials, stems from the early idea\cite{Hicks1993,Hicks1993a,Hicks1993b} (still controversial\cite{Mahan1996} and not clearly verified in experiments so far) that 2D electron confinement could burst the already large thermoelectric power of the SrTiO$_3$ bulk, with room-$T$ Seebeck coefficient ($S$) of several hundredths $\mu$V/K in the low carrier density range $n\sim$10$^{19}$ cm$^{-3}$,\cite{Okuda2001,Ohta2005}. In addition, with respect to the known Bi- and Te-based thermoelectric materials, oxides have the important advantage of being non toxic and structurally compatible for integration in multifunctional heterostructures. 

Remarkably large thermopower was reported for several 2D oxides. Most notably, an $S$ up to $\sim$ 1000 $\mu$V/K, with a strong dependence on the carrier density was measured in SrTiO$_3$/SrTi$_{0.8}$Nb$_{0.2}$O$_3$ superlattices\cite{Mune2007,Ohta2007,Ohta2007a,Ohta2008,Ohta2008a,Ohta2008b}. For what concerns LaAlO$_3$/SrTiO$_3$, the first measurement of Seebeck coefficient in the range from $T$=77 K to room-$T$ \cite{Pallecchi2010} gave values similar to those observed for low-doped SrTiO$_3$ bulk, but also revealed the possibility of tuning $S$ by a gate voltage, as previously highlighted in the implementation of SrTiO$_3$-based transistors.\cite{Ohta2009} An anomalous low-$T$ behavior of thermopower in LaAlO$_3$/SrTiO$_3$ was also reported.\cite{Lerer2011} The first $S$ measurement in the extended $T$= 4-300 K range \cite{Filippetti2012} revealed the presence of an impressively narrow and deep peak ($\sim$ 500 $\mu$V/K) below $T$=50 K, in striking contrast with the underlying diffusive regime which dictates a smooth and almost linear approach to zero with decreasing temperature. For assonance with the thermoelectric behavior of semiconductor-based quantum wells\cite{Fletcher1994,Fletcher2000,Fletcher2002}, it was logical to attribute the peak to the phonon-drag effect, i.e. the extra contribution to the electronic thermopower due to the drag of the electrons with the diffusing phonons, induced by the electron-phonon coupling. The presence of large phonon-drag reappeared later in ion-gated SrTiO$_3$\cite{Shimizu2015} and in strongly charge-depleted LaAlO$_3$/SrTiO$_3$\cite{Pallecchi2015}, the latter in form of spectacular Seebeck oscillations observed under negative gate field, with a Seebeck amplitude reaching record-high values of several tens mV/K, and attributed to the presence of localized states below the mobility edge.

In order to rationalize the general behavior of phonon-drag and its sensitivity to the quantum confinement and localization, here we use a combination of experiments and modeling to perform a comparative analysis of phonon-drag in several LaAlO$_3$/SrTiO$_3$ interfaces and doped SrTiO$_3$ crystals. We furnish a clear evidence that for the interfaces the low-$T$ thermopower is dominated by a marked phonon-drag peak, which instead is absent (or barely detectable) in the bulk. Furthermore, we demonstrate that the phonon-drag peak is a consequence of the tight 2D electron confinement typical of oxide heterostructures. Indeed, in the low-temperature limit, the coupling of acoustic phonons with 2D-confined electrons is enhanced by the loss of the crystal momentum conservation in the interface-orthogonal direction, enabling the interaction of the electron gas with many more phonon frequencies. The close relation between strong electron-acoustic phonon scattering and large phonon-drag was previously analyzed in Al$_x$Ga$_{1-x}$As/GaAs\cite{Tsaousidou2001}, MgZnO/ZnO\cite{Tsaousidou2014} and carbon nanotubes\cite{Tsaousidou2010}. For LaAlO$_3$/SrTiO$_3$, the huge phonon-drag peak can be understood as another manifestation of the strong electron-phonon coupling recently revealed by the polaronic nature of the 2DES\cite{Cancellieri2016}, and also proposed to be the source of its superconducting behavior\cite{Rosenstein}.

The article is organized as it follows: In Section \ref{exp_set} we describe the experimental setup; section \ref{res} is devoted to the description of results, separated in thermoelectric and transport measurements (\ref{expt_thermo}), measurements under field-effect (\ref{expt_fe}), and theory results (\ref{model}). In section \ref{concl} we draw our conclusions. Finally in the Appendices we describe the model used for our calculations in detail.

\section{Experimental setup}
\label{exp_set}

LaAlO$_3$/SrTiO$_3$ interfaces are prepared by pulsed laser deposition. On a TiO$_2$-terminated (001)-oriented SrTiO$_3$ crystal, 5 unit cells (uc) (sample B) and 10 uc (samples A and C) of LaAlO$_3$ are grown at substrate temperature of 650$^o$ C (sample A) and 800$^o$ C (sample B and C) in an oxygen pressure of 10$^{-5}$ mbar, and then annealed in an oxygen pressure of 200 mbar for one hour at 520$^o$ C before cooling down to room temperature in the same oxygen atmosphere.\cite{Cancellieri2010} A gold pad is evaporated on the back of the 0.5 mm thick substrate and used as a gate electrode, for field effect experiments. Commercially available SrTi$_{1-x}$Nb$_x$O$_3$ single crystals with different Nb doping are also measured. Seebeck coefficient is measured in a home-made cryostat, from 4K to room temperature, using an a.c. technique\cite{Calzona1993}. A sinusoidal heating power with a period of 150 s is supplied to one side of the sample, producing a thermal gradient of 0.15 K/mm. Hall effect and resistivity data are measured in a PPMS system by Quantum Design, from 4K to room temperature in magnetic fields up to 9 Tesla.

\section{Results}
\label{res}

\subsection{Thermoelectric and transport measurements}
\label{expt_thermo}

\begin{figure}
\centering
\includegraphics[clip,width=9cm]{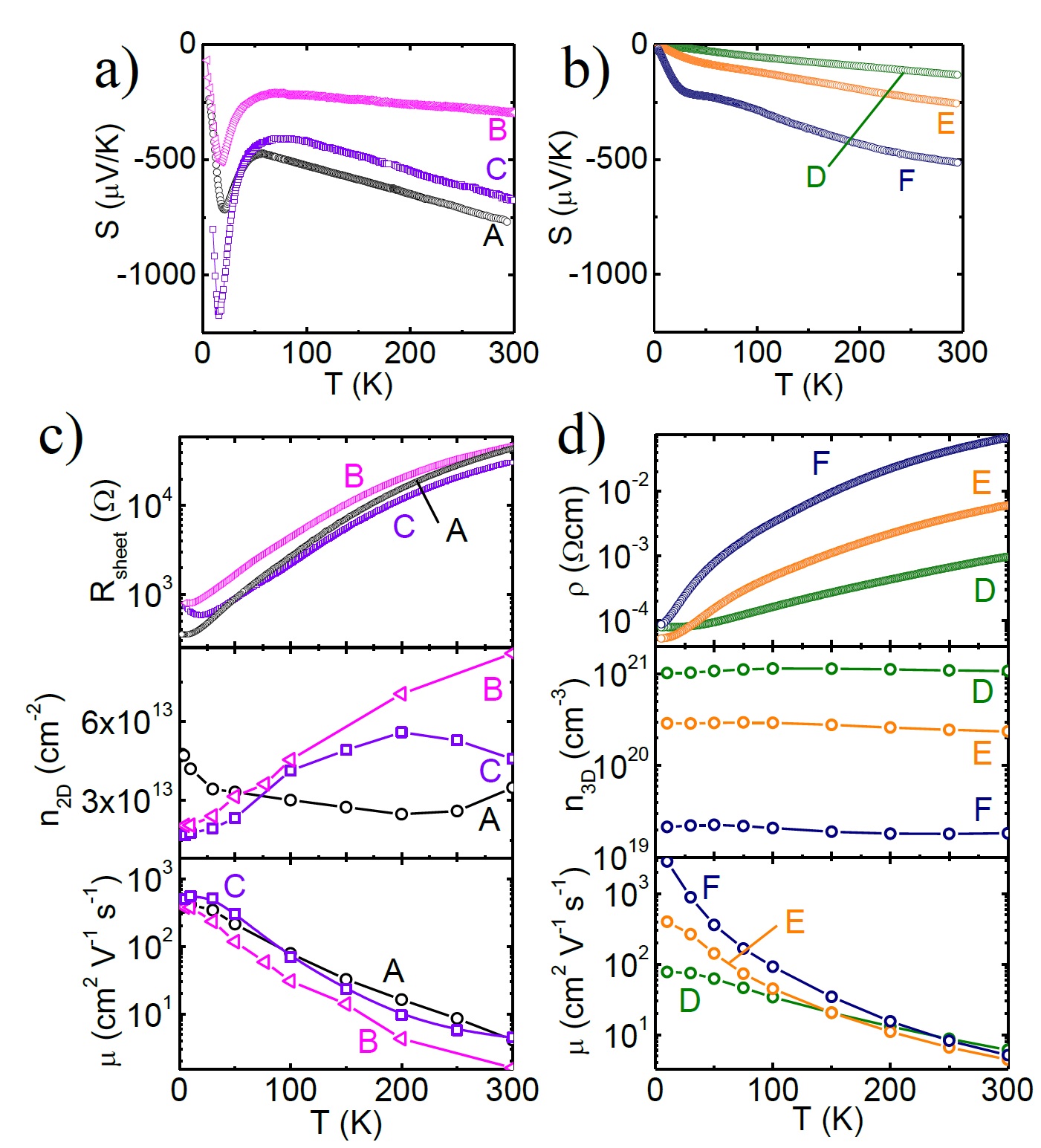}
\caption{
a): Seebeck coefficient $S$ measured for the three interfaces as a function of temperature. b): Seebeck coefficient for the three Nb-doped single crystals. c): sheet resistance ($R_{sheet}$), 2D carrier density ($n_{2D}$) and Hall mobility ($\mu$) for the interfaces. d): resistivity ($\rho$), 3D carrier density ($n_{3D}$) and Hall mobility ($\mu$) for the Nb-doped SrTiO$_3$ single crystals.
}
\label{exp}
\end{figure}

Figure \ref{exp}a) displays the Seebeck coefficient measured as a function of $T$ for the three different interface samples. We see that above 50-70 K, $S$ is linear for all the samples, as expected in the diffusive regime. The linear slopes, varying between -0.37 and -1.28 $\mu$V/K$^2$, are reported in Table \ref{tab1}, together with other measured transport quantities of the interfaces. Below 50 K a sharp peak associated to the phonon-drag mechanism is observed. The temperature position of the peak is around 15-20 K in all the cases, consistently with the fact that the maximum amplitude of the electron coupling with acoustic phonons is expected around $\theta_D$/10, if $\theta_D$ is the Debye temperature (for undoped SrTiO$_3$ ${\theta}_D$ $\sim$ 513 K\cite{Ahrens2007}). At higher $T$, on the other hand, other phonon scattering processes (namely boundary scattering, phonon-impurity, and phonon-phonon scattering) become dominant over electron-phonon, and phonon-drag rapidly fades.
If the peak position is substantially the same in all the samples, its amplitude is clearly sample-dependent (see Table \ref{tab1}), with a trend not trivially related to other fundamental characteristics such as carrier density, sheet resistance, or mobility (also reported in Table \ref{tab1}): at low $T$, $n_{2D}$ is rather similar for interfaces B and C, and a factor-two higher for A; consistently, $R_{\rm sheet}$ is nearly double for the formers. Nevertheless, the highest phonon-drag peak occurs in C ($|S|$ $\sim$1180 $\mu$V/K) and the smallest in B ($|S|$$\sim$500 $\mu$V/K), while A remains in the middle ($|S|$$\sim$715 $\mu$V/K). The Hall mobility is not quite helpful either in rationalizing the phonon-drag behavior: since a large phonon scattering due to defects or disorder plays in favor of phonon-drag suppression,\cite{Kaiser1987,Chatterjee1998} it could be reasonable to expect a relation between low-$T$ mobility and phonon-drag. However, our data in Table \ref{tab1} defy such a simple interpretation: samples A and B have same mobility but quite a different drag peak. With the help of modeling, we will see later that the difference in amplitude can be actually traced back to different 2DES confinement thickness and planar charge localization, features not easily determined experimentally. 
The other transport quantities shown in Fig. \ref{exp}c follow the expected behaviors: $R_{\rm sheet}$ increases with temperature and saturates to a residual, slightly sample-dependent value; the Hall-measured $n_{2D}$ for all samples stays within the usual 10$^{13}$-10$^{14}$ cm$^{-2}$ range, and the Hall mobility decreases with increasing $T$.

In Fig. \ref{exp}b) we report the Seebeck coefficient measured for three Nb-doped SrTiO$_3$ bulk samples with different doping concentrations. Differently from the interfaces, they are essentially phonon-drag free, with the exception of the lowest-doped sample F, showing a minor deviation from the linear behavior in a form of a rounded shallow bump below 50 K. A barely visible inflection is also present for sample E. We can conclude that for bulk samples phonon-drag should be minor or smaller than the diffusive contribution. Notice also that the absence of phonon-drag in SrTiO$_3$ crystals cannot be merely attributed to the chemical doping, which would suppress the phonon relaxation time: for sample F the mobility at low-$T$ is much higher than for the interfaces, thus we may argue that for this sample the electron-phonon vs. impurity scattering ratio is larger than for the interfaces, and yet, there is no significant drag peak. Finally, in Fig. \ref{exp}d) the transport properties of the bulk samples are shown as a function of temperature. The resistivity curves can be phenomenologically described by the Bloch-Grüneisen law\cite{Gruneisen1933} plus a T$^2$ term \cite{vandermarel2011} . The volume carrier densities n$_{3D}$ are almost constant in the whole temperature range. For all the three samples, n$_{3D}$ is about three times as large as the nominal doping indicating that some oxygen vacancies contribute in providing additional carriers. The mobility vs temperature curves are similar to those seen for the SrTiO$_3$/LaAlO$_3$ interfaces; the largest low-temperature value is reached for the least doped compound (2870 cm$^2$V$^-1$s$^-1$ for sample F). The relevant transport parameters of the three single crystals are summarized in Table \ref{tab1}.

\begin{table}[h]
\caption{\label{tab1} 
Some relevant transport parameters for our examined samples. From left to right: the $S$ slope vs. $T$ in the linear (diffusive) regime ($\mu$V/K$^2$); $S$ at room temperature ($\mu$V/K); the phonon-drag peak amplitude $S_{\rm peak}$ ($\mu$V/K); the Hall-measured carrier densities at $T$=4 K (in cm$^{-2}$ for the interfaces, cm$^{-3}$ for bulks); the sheet resistance $R_{\rm sheet}$ for the interfaces ($\Omega$) and the bulk resistivity $\rho$ for the bulks ($\Omega$cm) at $T$= 4 K; the Hall mobility at $T$= 4 K (cm$^2$/V/s).}
\begin{centering}
\begin{tabular}{l|c|c|c|c|c|c}
\hline
 LAO/STO & $dS/dT$ & $S$ & $S_{\rm peak}$ & $n_{2D}$ & $R_{\rm sheet}$ &  $\mu$  \\
\hline
A 	& -1.25	& -770	& -715 & 4.8$\times$10$^{13}$ & 350 & 380 \\
B 	& -0.37	& -290	& -500 & 2.0$\times$10$^{13}$ & 800 & 380 \\
C 	& -1.28	& -680	&-1180 & 1.7$\times$10$^{13}$ & 740 & 560 \\
\hline
\hline
 STO    & $dS/dT$ & $S$ & $S_{\rm peak}$ & $n_{3D}$ & $\rho$ &  $\mu$  \\
\hline
D 	& -0.40	& -134	&      & 1.0$\times$10$^{21}$ & 7.85$\times$10$^{-5}$ & 78 \\
E 	& -0.73	& -258	& -62  & 2.9$\times$10$^{20}$ & 5.25$\times$10$^{-5}$ & 400 \\
F 	& -1.20	& -517	& -210 & 2.2$\times$10$^{19}$ & 8.65$\times$10$^{-5}$ & 2870 \\
\hline
\end{tabular}
\end{centering}
\end{table}

\subsection{Field-effect measurements}
\label{expt_fe}

\begin{figure}
\centering
\includegraphics[clip,width=7cm]{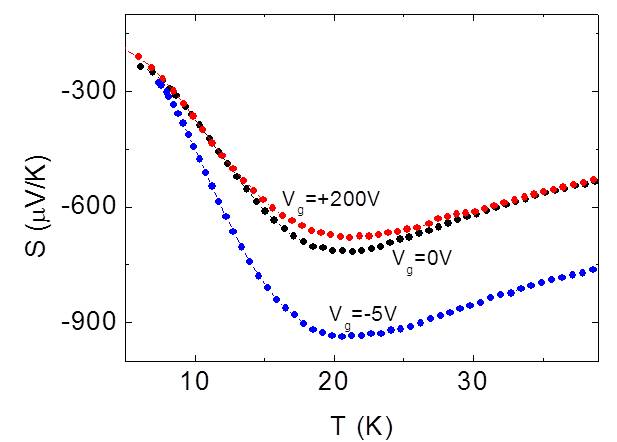}
\caption{Seebeck coefficient measured for interface sample A as a function of temperature for three values of the back gate voltage V$_g$.
}
\label{exp2}
\end{figure}

In Fig. \ref{exp2} we report Seebeck coefficient measurements for interface sample A under a back-gate voltage $V_g$ (no "poling" protocol" is performed here\cite{note1}). We see that phonon drag peak is significantly modulated by the gate voltage. The asymmetry of S with respect to the sign of V$_g$ is related to the non-linear field dependence of the dielectric permittivity, and disappears above 77K (see Ref.\cite{Pallecchi2010}) when permittivity does not depend anymore on the electric field. In the accumulation regime, $V_g$ can be varied up to +200 V without detectable leakage and the magnitude of $S$ at the phonon drag peak (S$_{\rm peak}$) is suppressed by only 5\%. On the other hand, negative gate voltages are very effective in depleting the interface of carriers. For $V_g$=-5V we obtain a peak enhancement of 30\%, while the diffusive regime at high temperature shows negligible modulation with the field, consistently with previous data\cite{Pallecchi2010}. The application of a negative $V_g$ is known to produce a series of outstanding effects on the 2DES characteristics, related to charge depletion and increase of electron confinement.\cite{Biscaras2012} In particular, Ref.\onlinecite{Pallecchi2010} shows that negative gate fields of order MV/m can reduce the 2DES thickness up to a factor-2 in the high temperature regime $T>$ 77 K. Thus, the large phonon-drag modulation by field effect is a clear evidence that phonon-drag is crucially related to the 2D extension of the gas.

\subsection{Phonon-drag modeling}
\label{model}

To shed light into the phonon-drag mechanism at the fundamental level, we used the theory first developed by Bailyn \cite{Bailyn1967} and then adapted to 2DES systems by Cantrell and Butcher\cite{Cantrell1986,Cantrell1987a,Cantrell1987b,Tsaousidou2001,Tsaousidou2010,Tsaousidou2014}, based on the Boltzmann Transport Equation (BTE)\cite{Allen1996} for coupled electrons and phonons. To make calculation affordable, we describe the 2DES electronic structure by an anisotropic effective mass modeling (previously used for a series of oxides\cite{Filippetti2012,Delugas2013,Delugas2013a,Delugas2015,Puggioni2012}) and the acoustic phonon frequencies by a simple linear dispersion. Hereafter we briefly sketch the final phonon-drag expression, leaving the detailed description to the Appendices. The phonon-drag is:

\begin{equation}
\begin{split}
S_j^{pd} =-\frac{e\nu_s^2}{\left(2\pi\right)^3 \sigma_j k_B T^2}\sum_n\sqrt{\frac{m_{nx}^*m_{ny}^*}{{m^*_{nj}}^2}}
\int\limits_{\epsilon_n^0}^{\epsilon_n^0+W_n} \!\!\!\!\! d\epsilon \,f(\epsilon)\,\tau_n(\epsilon) \\
\hspace{-1 cm} \times \int\limits_0^{q_0} \!\! dq_p \, q_p^3 \int\limits_{-q_0}^{q_0}\!\! dq_z 
\frac{N_q (1-f(\epsilon+\hbar\omega_q))} {\sqrt{C_{0,n}^2-X_{0,n}^2} }
\frac{A_n(q,q_z)} {\tau^{-1}_{ph}(q)}
\end{split}
\label{eq_pd}
\end{equation}

where $\nu_s$ is the speed of sound, $\sigma_j$ the 2D conductivity (in $\Omega^{-1}$ ), {\it m}$_{nj}^*$ the effective mass of band $n$ in direction $j$ ($x$ or $y$); $\epsilon$ is the electron energy, $f({\epsilon})$ and $N_q$ are electron and phonon occupancies, respectively, and $A_n(q,q_z)$ the electron-phonon coupling amplitude; $\tau_n$($\epsilon$) and $\tau_{ph}$($q$) are electron and phonon relaxation times, respectively; $\epsilon_n^0$ and $W_n$ are band bottom and bandwidth, $C_{0,n}$ and $X_{0,n}$ quantities dependending on phonon and electron energy (see Eq.\ref{eq_ap13} in Appendix I). The integral over phonon wavevector ${\bf q}$ is solved in cylindrical coordinates, $q_p$=$\sqrt{q_x^2+q_y^2}$ and $q_z$ are planar and orthogonal components, $q$=$\sqrt{q_p^2+q_z^2}$. In Eq.\ref{eq_pd} the crucial quantity which governs the phonon-drag magnitude is the ratio of electron-acoustic phonon coupling to phonon lifetime, with the former expressed as:

\begin{equation}
A_n(q,q_z) ={\tilde A}(q) F_n(q_z);\hspace{0.2 cm}F_n(q_z)=\left| \int\limits_t dz\> \psi^2_n(z)\> e^{iq_z z} \right|^2
\label{eq_pd2}
\end{equation}

where ${\tilde A}(q)$ includes the deformation potential and the piezoelectric contributions\cite{Tsaousidou2001}, and $\psi_n(z)$ is the wavefunction of the 2D-confined electrons, whose Fourier-transform $F_n$ governs the coupling of 3D phonons with the electrons confined along $z$ within a slice of thickness $t$. In the $t$ $\rightarrow$ $\infty$ limit $F_n$$\rightarrow$$\delta_{q_z,0}$ i.e. only zero-wavelength phonons can couple with electrons, and the coupling amplitude goes back to the 3D case, where ${\bf q}$ is fully determined by the crystal momentum conservation $\bf k$'=$\bf k$+$\bf q$; in case of “infinite” confinement ($t$=0), on the other hand, all $q_z$ up to the Debye wavelength do contribute to the coupling, and $F_n$=1. For what concerns $\tau_{ph}(q)$, we use the low-temperature modeling developed by Callaway\cite{Callaway1959} (see Eq.\ref{eq_ap16} in Appendix I), while band structure parameters {\it m}$_{nj}^*$, $\epsilon_n^0$, and $W_n$, are taken from our previous ab-initio results.\cite{Filippetti2012,Cancellieri2014}


In the calculation we want to put in evidence the key features which govern the phonon-drag amplitude: the gas thickness $t$, the in-plane localization (thus {\it m}$_{nj}^*$), and the $n_{2D}$ charge density. Those are all quite difficult to be precisely determined in the experiments, since crucially affected by a number of hard-to-control conditions, such as structural disorder, oxygen vacancies, cation intermixing. Exploiting the flexibility of band modeling, we thus treat them as variable parameters to explore the phonon-drag behavior in a range of different conditions. For what concern the 2DES thickness, ab-initio results\cite{Pentcheva2006,Park2006,Lee2008,Popovic2008, Pentcheva2008,Pentcheva2009,Janicka2009,Delugas2011,Stengel2011,Filippetti2012,Cancellieri2014} show that for the charge density of interest ($\sim$2-4$\times$10$^{13}$ cm$^{-2}$) the gas is entirely included in a few ($\sim$2,3) d$_{xy}$ states confined in the TiO$_2$ layers closest to the interface, while only above $\sim$ 6$\times$10$^{13}$ cm$^{-2}$ the more extended d$_{xz}$, d$_{yz}$ states sets in. However, in order to evaluate the scaling with respect to the 2D confinement, it is convenient to replace the actual squared wavefunctions with Gaussian envelope functions of variable thickness $t$ (see Appendix-I for details). In Fig. \ref{calc} we show three sets of phonon-drag calculations relative to different values of effective masses and $n_{2D}$, each of them for variable $t$ from a single unit cell up to 100 nm (i.e. in the bulk limit). Fig. \ref{calc}a shows the phonon-drag for masses derived from ab-initio calculations ($m_x^*$=$m_y^*$=0.7$m_e$ for d$_{xy}$ states) which represent the 'clean' interface limit, while Fig. \ref{calc}b is obtained using 'fattened' effective masses (1.4$m_e$ for d$_{xy}$) mimicking an increased in-plane charge localization, typically associated with structural distortions and disorder; finally, in Fig. \ref{calc}c we use 'ideal' masses and an increased $n_{2D}$. (notice that the explicit mass dependence in Eq.\ref{eq_pd} vanishes for isotropic masses, but the most crucial mass dependence is implicitly included in $f({\epsilon})$ through the Fermi energy). In Fig. \ref{calc} we also report diffusive Seebeck coefficient, sheet resistance, and mobility calculated by BTE for the different masses and densities, and $t$=1 nm.

\begin{figure}
\centering
\includegraphics[clip,width=9cm]{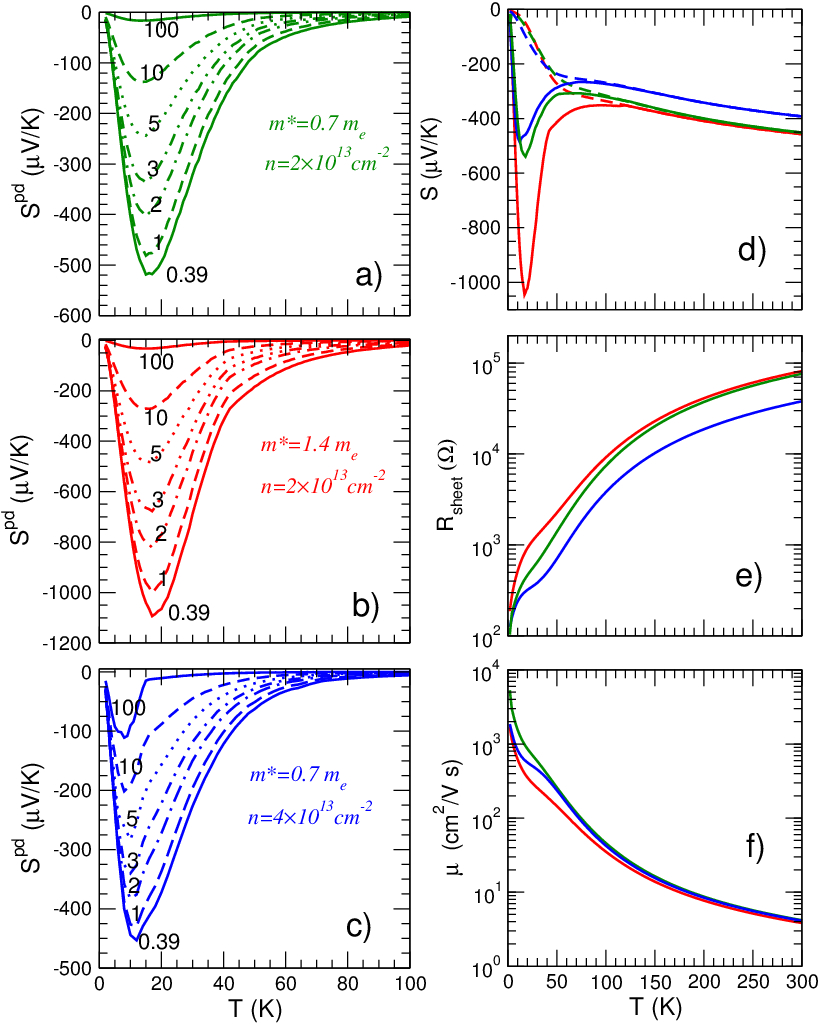}
\caption{
a), b), c): Model calculation of phonon-drag vs. temperature at varying gas thickness $t$ (indicated for each line in nm). Each panel corresponds to different values of effective masses and charge densities, also reported on each panel. d): diffusive Seebeck coefficient (dashed lines) and total (diffusive plus phonon-drag) Seebeck coefficient (full lines) for the three sets of calculations and $t$= 1 nm. Green curves are for case a), red curves for b) and blue for c). d) sheet resistance; e) electron mobility.
}
\label{calc}
\end{figure}

Overall, a good match with the observations is obtained. Our phonon-drag values qualitatively reproduce the major features observed in the experiment: the phonon-drag appears as a sharp peak centered at $T$$\sim$20 K, and for $T>$50 K it disappears behind the diffusive Seebeck coefficient. The peak amplitude crucially depends on both 2DES thickness and effective masses. In fact, we see that only for thickness of $\sim$10 nm or less the phonon-drag peak becomes significantly visible above the low-temperature diffusive background. In particular, assuming as maximum possible confinement the thickness of a single unit cell ($t$=0.39 nm) we obtain peaks higher than 500 $\mu$V/K and 1000 $\mu$V/K for the two sets of masses, which are in the range of the values observed for the interfaces. More complicated is tracing the phonon-drag dependence on the charge density: a larger charge implies higher Fermi energy but also more occupied bands (i.e. more bands actively contributing to the phonon-drag). The net result is $t$-dependent: for weakly confined 2DES, the second aspect prevails, and the phonon-drag increases with the density. In the narrow thickness limit, the two effects neary compensate and phonon-drag is slightly larger for the lower density. Overall, the effect of a pure charge-filling on the phonon-drag appears to be weaker than confinement and localization. Clearly, in actual cases these three ingredients (thickness, localization, and density) are tightly interlaced with each other, thus a fine-tuned interpretation of the differences observed among the interface samples is cumbersome. What is significant in the simulation, on the other hand, is that, assuming realistic values for $n_{2D}$, the phonon-drag scaling with thickness and effective masses spans the whole range of measured values, from the bulk to the tightly-confined 2DES regime.

\section{Conclusions}
\label{concl}

In conclusion, we reported the measured thermopower for several LaAlO$_3$/SrTiO$_3$ interfaces, in comparison with that of Nb-doped SrTiO$_3$ crystal samples. We give evidence that for the interfaces, thermopower below 50 K is dominated by a high phonon-drag peak, reaching huge values of order $\sim$mV/K. On the other hand, phonon-drag is substantially absent, or barely apparent, in bulks. With the help of numerical modeling, we traced back the presence of such a large phonon-drag peak to the tight 2D confinement of the 2DES and to the charge localization at the interface: specifically, the 2D thickness establishes the cut-off on the wavelength of the acoustic phonons which can couple to the confined electrons, while planar localization crucially controls the Fermi energy and in turn the phonon-drag amplitude. From our analysis, phonon-drag emerges as a remarkable “marker” of charge localization and confinement in 2D heterostructures, and a significant parameter of classification for 2DES systems in general.

\begin{acknowledgments}
Work supported by Italian MIUR through projects FIRB RBAP115AYN ‘Oxides at the nanoscale, multifunctionality and applications’; PRIN 2010NR4MXA ‘OXIDE’; the Swiss National Science Foundation through the ‘Thermoelectric oxides TEO’ project and Division II and the European Research Council under the European Union’s Seventh Framework Programme (FP7/2007–2013)/ERC Grant Agreement No. 319286 (Q-MAC). A.F. acknowledges computational support by CRS4 Computing Center (Piscina Manna, Pula, Italy).
\end{acknowledgments}


\section*{Appendix I - Phonon-drag modeling in 2D}
\label{ap1}

We follow the original Cantrell-Butcher (CB) formulation\cite{Cantrell1987a,Cantrell1987b} for the expression of phonon-drag in 2D heterostructures. The theory is developed starting from the Boltzmann Transport Equation within relaxation time approximation for a coupled system of 3D acoustic phonons and 2D electrons. Using the same CB notations, assuming intra-band scattering only, the phonon drag in direction $j$=($x$, $y$) can be expressed as:

\begin{eqnarray}
S_j^{pd}=\left({\frac{2e}{\sigma_j A k_B T^2}}\right) \sum_{nk,nk',q}\hbar\omega_q
\left( {\Gamma_{nk,nk'}({\bf q})\over\tau^{-1}_{ph}({\bf q}) }\right) \nonumber \\ 
\times v_j({\bf q}) \left[ \tau(n{\bf k}) v_j(n{\bf k})-\tau(n{\bf k}') v_j(n{\bf k}') \right]
\label{eq_ap1}
\end{eqnarray}

with $A$ the unit area, $\sigma_j$ the 2D conductivity (in $\Omega^{-1}$), ${\bf q}$ = ($q_x$, $q_y$, $q_z$) and $k$ = ($k_x$, $k_y$) phonon and electron crystalline momenta, $\tau_{ph}({\bf q})$ and  $\tau(n{\bf k})$ phonon and electron relaxation times, $v_j({\bf q})$ and $v_j(n{\bf k})$ phonon and electron velocities, and $\Gamma_{nk,nk'}({\bf q})$  the electron-acoustic phonon scattering rate:

\begin{eqnarray}
 \Gamma_{nk,nk'}({\bf q}) = f_{n{\bf k}}\left(1- f_{n{\bf k'}}\right) N_{\bf q} A_n({\bf q}) 
\nonumber \\
\times \delta\left( \epsilon_{n{\bf k}'}-\epsilon_{n{\bf k}}-\hbar\omega_{\bf q} \right) \delta_{\bk',\bk+\bq_p}
\label{eq_ap2}
\end{eqnarray}

where $f_{n\bk}$ and $N_{\bq}$ are electron and phonon occupancies, $A_n(\bq)$ is the coupling amplitude, and the two delta functions account for energy and in-plane momentum conservation ($\bq_p$=($q_x$, $q_y$)). Notice that Eq.\ref{eq_ap2} is written for the absorption process, but emission is implicitly accounted for in Eq.\ref{eq_ap1}. In Ref.\onlinecite{Cantrell1987b} Eq.\ref{eq_ap1} is made treatable assuming linear phonon dispersion and parabolic band modeling for electrons. Here we follow the same strategy, except for the generalization to anisotropic band masses in the plane, which is better suited to describe the t$_{2g}$ states of the 2DEG. It is easy to see that:

\begin{eqnarray}
v_j(\bq) \left[ \tau_{n\bk}v_j(n\bk)-\tau(n\bk')v_j(n\bk') \right]
=-{\hbar v_s \over m_{nj}^* } {q_j^2\over q}\tau(n\bk) \nonumber \\
\label{eq_ap3}
\end{eqnarray}

where $m_{nj}^*$ is the effective mass of $n^{th}$ band in direction $j$; also for simplicity we assume $\tau(n\bk)=\tau(n\bk')$. Notice that the minus sign in Eq.\ref{eq_ap3} comes from the fact that for positive band curvature (electrons) the band velocity increases with $\bk$; it follows that phonon drag is negative for electrons, just like diffusive thermopower. In order to treat anisotropic bands we introduce the following change of in-plane $j$=($x$, $y$) variables:

\begin{eqnarray}
k_j=K_j \sqrt{ m_{nj}^* \over m };\hspace {0.5cm} q_j=Q_j \sqrt{ m_{nj}^* \over m }
\label{eq_ap4}
\end{eqnarray}

where $m$ is an auxiliary mass which, using in-plane momentum conservation, allows to write:

\begin{eqnarray}
\epsilon_{n\bk}= { \hbar^2 K^2 \over 2m};\hspace {0.5cm} \epsilon_{n,\bk+\bq_p}=\epsilon_{n,\bK+\bQ}= { \hbar^2 (\bK+\bQ)^2 \over 2m}
\label{eq_ap5}
\end{eqnarray}

\begin{eqnarray}
\delta\left( \epsilon_{\bK+\bQ}-\epsilon_{\bK}-\hbar\omega_{\bq} \right)=
\delta\left( {\hbar^2 Q^2 \over 2m} + {\hbar^2 KQ \,\cos\theta \over m} -\hbar\omega_{\bq} \right)
\nonumber \\
\label{eq_ap6}
\end{eqnarray}

where $\theta$ is the angle formed by $\bK$ and $\bQ$. It is convenient to introduce another variable change:

\begin{eqnarray}
X={\hbar^2 KQ \over m}\cos\theta=C_0 \,\cos\theta; \hspace{0.5cm}X_0
=\hbar\omega_q -{\hbar^2 Q^2 \over 2m}
\label{eq_ap7}
\end{eqnarray}

Then, we solve the sum over $\bK$ in 2D for a fixed $\bK$ in radial coordinates, taking the azimuth angle to be that formed by $\bK$ and $\bQ$. It is easy to see that:

\begin{eqnarray}
d\theta=-dX{1\over \sqrt{C_0^2-X^2}};\hspace{0.5cm}KdK={m\over\hbar^2}\,d\epsilon_K
\label{eq_ap8}
\end{eqnarray}

\begin{eqnarray}
\sum_k \delta\left( \epsilon_{\bK+\bQ}-\epsilon_{\bK}-\hbar\omega_{\bq} \right)=
{A\over (2\pi)^2\hbar^2}\sqrt{m_{nx}^*m_{ny}^*} \nonumber \\
\times \int d\epsilon\!\!\! \int\limits_{X\in[-C_0,C_0]}dX
{\delta(X-X_0) \over \sqrt{C_0^2-X^2}}
\label{eq_ap9}
\end{eqnarray}

The argument of the square root is always positive, since $X\in[-C_0,C_0]$; to have a non-vanishing integral in $dX$ it must be -$C_0\leq X_0 \leq C_0$, i.e.:

\begin{eqnarray}
{\hbar^2 (\bK-\bQ)^2\over 2m}\leq \epsilon_{n\bk}+\hbar\omega_{\bq} \leq {\hbar^2 (\bK+\bQ)^2\over 2m}
\label{eq_ap10}
\end{eqnarray}

which has a simple interpretation: after absorption the carrier energy must be higher (lower) than the band energy corresponding to antiparallel (parallel) $\bK$ and $\bQ$ orientation. We have:

\begin{eqnarray}
\int\limits_{X\in[-C_0,C_0]}dX{\delta(X-X_0) \over \sqrt{C_0^2-X^2}}
={2\over \sqrt{C_0^2-X_0^2}}
\label{eq_ap11}
\end{eqnarray}

Since $\theta\in[0,2\pi]$ there are always two values ($\cos$($\theta_0$) and $\cos$(-$\theta_0$)) for which the delta function is non-vanishing. Inserting Eq.\ref{eq_ap2} in Eq.\ref{eq_ap1}, solving for the energy-conserving delta function according to Eqs.\ref{eq_ap9} and \ref{eq_ap11}, and using $\omega_{\bq}$ = $\nu_s q$, the phonon drag in 2D becomes:

\begin{eqnarray}
S_j^{pd}=-{4e v_s^2 V \over (2\pi)^5 \sigma_j k_B T^2}\sum\limits_n\sqrt{m_{nx}^*m_{ny}^*\over (m_{nj}^*)^2}\int d^3q \> q_j^2 \,{N_{\bq} A_n(\bq) \over \tau_{ph}^{-1}(\bq)} \nonumber \\
\times \int\limits_{\epsilon_n^0}^{\epsilon_n^0+W_n} \!\!\!\! d\epsilon\> f_{\epsilon} \left(1- f_{\epsilon+\hbar\omega_q}\right) \tau_n(\epsilon){1\over \sqrt{C_{0,n}^2-X_{0,n}^2}}\hspace{0.8cm}
\label{eq_ap12}
\end{eqnarray}

\begin{eqnarray}
C_{0,n}^2={2\hbar^2 {\tilde{q}_n}^2\over m_{nx}^*m_{ny}^*}\epsilon_K;\hspace{0.3cm}
X_{0,n}=\hbar\omega_q-{\hbar^2{\tilde{q}_n}^2\over 2 m_{nx}^*m_{ny}^*}
\label{eq_ap13}
\end{eqnarray}

where ${\tilde{q}_n}^2=q_x^2 \,m_{ny}^*+ q_y^2 \,m_{nx}^*$. In 2D the electron-acoustic phonon scattering amplitude can be written, at the simplest level of approximation:\cite{Ridley1999}

\begin{eqnarray}
A_n(q,q_z)=\left[ C_{DP}\,q+C_{PZ}{q^3\over (q^2+q_D^2)^2} \right] F_n(q_z);
\label{eq_ap14}
\end{eqnarray}

\begin{eqnarray}
C_{DP}={\pi D^2\over V\rho\nu_s};\hspace{0.5cm}C_{PZ}={\pi e^2 \nu_s K_{em}^2 \over V k_0 k};\\
F_n(q_z)=\left| \int\limits_t dz\> \psi^2_n(z)\>e^{iq_z z}\right|^2
\label{eq_ap15}
\end{eqnarray}

$C_{DP}$ and $C_{PZ}$ account for deformation potential and piezoelectric scattering, respectively; $D$ is the deformation potential, $\rho$ the mass density, $k_0$ and $k$ vacuum permittivity and dielectric constant, $K_{em}$ the 3D-averaged electromechanical coupling, and $q_D$ the Debye screening length; $\psi_n(z)$ is the space-localized wavefunction of the scattered electrons. The deformation potential term describes the coupling of electrons with longitudinal acoustic waves treated as an homogeneous strain. The piezoelectric scattering is the additional contribution due to the coupling with the electric field produced by the strain. For a non-polar system ($K_{em}$ =0) or in the limit of large doping concentration (i.e. strong Debye screening) the second term vanishes and only the deformation potential contributes to the acoustic scattering. In case of small screening ($q_D$=0) and highly ionic compounds, on the other hand, the piezoelectric contribution ($\sim$1/q) may become dominant at small $q$. Assuming a very long Debye screening length ($q_D$ $\sim$0) screening becomes discardable, and the electron-phonon scattering increases ,\cite{Fletcher2002} i.e. in the low-density charge-localized limit, carrier mobility is so small that the piezoelectric interaction becomes unscreened. In the case of LaAlO$_3$/SrTiO$_3$ both terms are relevant and should be included in the treatment.

\begin{figure}
\centering
\includegraphics[clip,width=9cm]{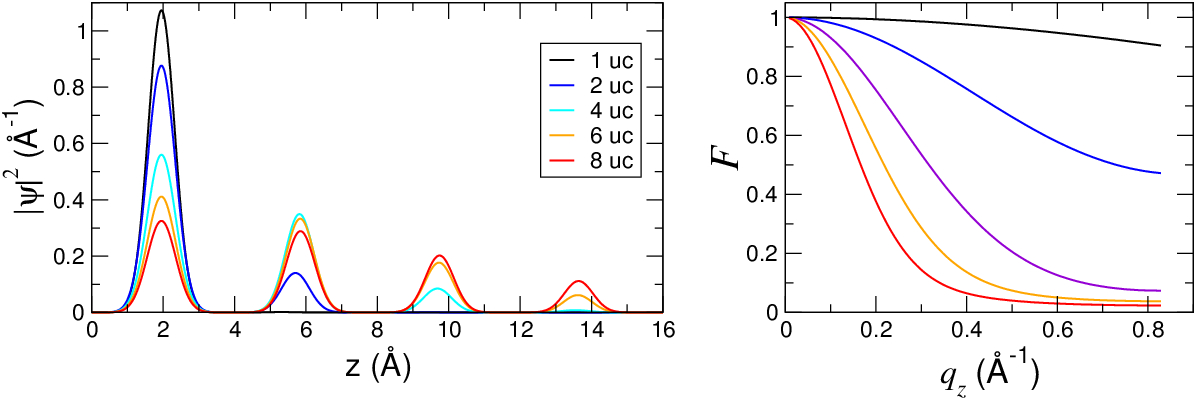}
\caption{
Left panel: $z$-localized squared wavefunctions with varying localization length, normalized to unity. Right: corresponding Fourier transforms. Each wavefunction is built as a sum of Gaussian functions, centered in the middle of a perovskite unit cell, further interpolated by a Gaussian envelop of varying thickness.
}
\label{ff}
\end{figure}

In the form factor $F_n$ we assume the same initial and final wavefunction for the scattered carrier; this is reasonable since $F_n$ depends weakly on the specific wavefunction shape; what matters the most is its overall extension, i.e. the thickness $t$ of the well. In Fig.\ref{ff} we show the relation between a series of normalized squared wavefunctions obtained by a superposition of Gaussian functions, with localization length progressively increased from one to eight unit cells, and the corresponding Fourier transforms $F_n$: in case of maximum localization, all the charge is enclosed in a single unit cell (black line), and the corresponding $F_n$ weight is about unity, i.e. all the phonons with $q_z$ up to the Debye wavelength ($q_z$ $\sim$ 0.83 \AA$^{-1}$ in SrTiO$_3$) contribute to the scattering; for a squared wavefunction localized in 8 u.c. (about 31 \AA$\,$ in SrTiO$_3$) on the other hand, only phonons with $q_z$ $>$ 0.2 \AA$^{-1}$ do contribute. In the limit of complete delocalization ($t$ $\rightarrow$ $\infty$) is $F_n$ $\rightarrow$ $\delta(q_z,0)$. Thus, the 2DEG confinement is a major factor of phonon-drag amplification in oxide heterostructures, as discussed in the main text. 

Finally, for the phonon relaxation time we use the Callaway formula:\cite{Callaway1959}

\begin{eqnarray}
\tau_{ph}^{-1}= A\,\omega_q^4 + B\,T^3 \omega_q^2+{\nu_s\over L}
\label{eq_ap16}
\end{eqnarray}

including scattering by point impurities (first term), phonon-phonon scattering (second term), and boundary scattering (third term; here $L$ is a characteristic sample length; in our model $L$=1 mm). In our calculations, $A$ and $B$ are adjusted to have a phonon-drag value in the bulk limit ($t$ $\rightarrow$ $\infty$) which is small with respect to the diffusive Seebeck, consistently with what is observed in the experiments; then these values are kept fixed while $t$ is progressively reduced, in order to describe the scaling effect purely due to the confinement thickness on the phonon-drag amplitude.

\section*{Appendix II - Approximate solutions for the phonon-drag integration in 2D}
\label{ap2}

Using the series of Eqs. from \ref{eq_ap12} to \ref{eq_ap16}, phonon drag can be obtained by numerical integration, and evaluated at any given temperature and doping; however this requires the integration over 4 coupled coordinates (3 for the phonon wavevector and one for the electron energy). To reduce the computational weight, we can follow two routes:

a) {\it Isotropic 2D approximation}: in plane we take $q_j^2$ $\sim$ 1/2 $q_p^2$;  it follows that the arguments of the square root in Eq.\ref{eq_ap11} become dependent on 2 coordinates only, and the integral over q can be expressed in cylindrical coordinates. Thus phonon drag is reduced to a 3-variable integration:

\begin{eqnarray}
S_j^{pd}=-{e\nu_s^2 \over (2\pi)^3 \sigma_j k_B T^2}
\sum\limits_n\sqrt{m_{nx}^*m_{ny}^*\over (m_{nj}^*)^2}
\int\limits_{\epsilon_n^0}^{\epsilon_n^0+W_n} \!\!\!d\epsilon \,f_{\epsilon}\,\tau_n(\epsilon) \nonumber \\
\times \int\limits_0^{q_0} dq_p\> q_p^3 \int\limits_{-q_0}^{q_0}dq_z\> {N_q\over \tau_{ph}^{-1}(q)}{(1- f_{\epsilon+\hbar\omega_q})\over \sqrt{C_{0,n}^2-X_{0,n}^2}} \nonumber \\
\times \left[C_{DP}\,q+C_{PZ}{q^3 \over (q^2+q_D^2)^2} \right] F_n(q_z) 
\nonumber \\
\label{eq_ap20} 
\end{eqnarray}

which corresponds to Eq.\ref{eq_pd} and was actually used for our calculations. Here $\pi q_0^2$=$(2\pi)^2/A$, and $q_0$=$2\sqrt{\pi}/a_0$. Also, electron and phonon energies must obey the constraints:

\begin{eqnarray}
\hbar\omega_q\leq{\hbar^2 q_p^2 \over 2\tilde{m}}+\hbar q_p \sqrt{2\epsilon_{nk}\over\tilde{m}}
\end{eqnarray}

\begin{eqnarray}
\hbar\omega_q \geq {\hbar^2 q_p^2\over 2\tilde{m}}-\hbar q_p \sqrt{2\epsilon_{nk}\over\tilde{m}}
\end{eqnarray}

with
\begin{eqnarray}
\tilde{m}_n={2 m_{nx}^* m_{ny}^* \over m_{nx}^*+ m_{ny}^* }
\end{eqnarray}

b) {\it small phonon frequencies}. In case of small phonon energies and low T, the following Equation holds (adopted in the CB work):

\begin{eqnarray}
f_{\epsilon}\left(1- f_{\epsilon+\hbar\omega_q}\right) \sim {\hbar\omega_q\over 1-e^{-{\hbar\omega_q\over k_B T}} } =\hbar\omega_q (N_q+1)\delta(\epsilon-\epsilon_F)\nonumber \\
\end{eqnarray}

thus in Eq.\ref{eq_ap12} the electron energy integral can be factorized and solved, and the phonon-drag becomes only dependent on phonon coordinates: 

\begin{eqnarray}
S_j^{pd}=-{2e\nu_s^2 \over (2\pi)^4 \sigma_j k_B T^2}
\sum\limits_n\sqrt{m_{nx}^*m_{ny}^* \over (m_{nj}^*)^2} \tau_n(\epsilon_F) \nonumber \\
\times \int d^3q \> q_j^2 \left( {\hbar\omega_q N_q(N_q+1) \over \tau_{ph}^{-1}(q) \sqrt{C_{0,n}^2(\epsilon_F)-X_{0,n}^2} } \right) \nonumber \\
\times \left[ C_{DP}\,q+C_{PZ}{q^3\over (q^2+q_D^2)^2}\right] F_n(q_z)
\label{eq_ap25}
\end{eqnarray}

with:

\begin{eqnarray}
C_{0,n}^2(\epsilon_F)={2\hbar^2 {\tilde{q}_n}^2\over m_{nx}^*m_{ny}^*}\epsilon_F;\hspace{0.5cm}
X_{0,n}=\hbar\omega_q-{\hbar^2{\tilde{q}_n}^2\over 2 m_{nx}^*m_{ny}^*}\nonumber \\
\end{eqnarray}

The approaches a) and b) give results to within 10-15\% from each other in the fully converged limit of electron energy and phonon wavevector integration. The results shown in Fig.\ref{calc} are obtained using Eq.\ref{eq_ap20}.

\section*{Appendix III - Values of parameters used for the calculation}
\label{ap3}

Several parameters entering the phonon-drag modeling are unknown or difficult to calculate for LaAlO$_3$/SrTiO$_3$; in this case we rely on values appropriate for SrTiO$_3$. For the phonon drag we use $\nu_s$=7.9$\times$10$^5$ cm/s, $k$=300, $D$=8 eV, $K_{em}$=0.35; for the phonon relaxation time $A$=0.5$\times$10$^{-41}$ (adimensional), $B$=0.5$\times$10$^{-20}$  K$^{-3}$, and $L$=0.1 mm. The electronic band structure is described by a multiband effective mass modeling, including three t$_{2g}$ bands for each unit cell, with masses  ($m_x^*$, $m_y^*$)= (0.7, 0.7)$m_e$ for $d_{xy}$, (0.7, 8.8)$m_e$ for $d_{xz}$, (8.8, 0.7)$m_e$ for $d_{yz}$; the band bottoms $\epsilon_n^0$ are scaled according to the ab-initio band structure calculations at varying charge density\cite{Cancellieri2014}: for the $d_{xy}$ states $\epsilon_1^0$=0, $\epsilon_2^0$=20 meV, $\epsilon_{i+1}^0$ =30 meV, for $i$=2,3,4,... ($i$ is the number of u.c. distance from the interface); for the  $d_{xz}$ and $d_{yz}$ states we take $\epsilon_i^0$=30 meV for each $i$, since they are substantially spread within the substrate and their confinement along $z$ is discardable for charge densities $n_{2D}$ $\sim$ 10$^{13}$ cm$^{-2}$. Finally, for the calculation of diffusive Seebeck and conductivity we use the same formulation and parameters previously adopted in Refs.\onlinecite{Filippetti2012} and \onlinecite{Pallecchi2015}.


%


\end{document}